\documentclass[conference]{IEEEtran}

\usepackage{amsmath}
\usepackage{amsfonts}
\usepackage{graphicx}
\usepackage[short]{optidef}
\usepackage[margin=0.70in]{geometry}
\usepackage{cite}
\providecommand{\abs}[1]{\lvert#1\rvert} 

\IEEEoverridecommandlockouts
\begin{document}

\title{Cyclic Prefix Direct Sequence Spread Spectrum Capacity Analysis}
\date{1 October 2019}
\author{%
    \IEEEauthorblockN{Brent A. Kenney\IEEEauthorrefmark{1}, Stephen N. Jenkins\IEEEauthorrefmark{1}, Arslan J. Majid\IEEEauthorrefmark{2}, Hussein Moradi\IEEEauthorrefmark{2}, and Behrouz Farhang-Boroujeny\IEEEauthorrefmark{1}}
    \IEEEauthorblockA{\IEEEauthorrefmark{1}Electrical and Computer Engineering Department, University of Utah, Salt Lake City, Utah, USA}
    \IEEEauthorblockA{\IEEEauthorrefmark{2}Idaho National Laboratory, Salt Lake City, Utah, USA\\}
\thanks{This manuscript has in part been authored by Battelle Energy Alliance, LLC under Contract No. DE-AC07-05ID14517 with the U.S. Department of Energy. The United States Government retains and the publisher, by accepting the paper for publication, acknowledges that the United States Government retains a nonexclusive, paid-up, irrevocable, world-wide license to publish or reproduce the published form of this manuscript, or allow others to do so, for United States Government purposes. \bf{STI Number: INL/CON-19-56283}.}
}

\maketitle
\begin{abstract}
Cyclic Prefix Direct Sequence Spread Spectrum (CP-DSSS) is a novel waveform that is proposed as a solution to massive machine type communications (mMTC) for 5G and beyond.  This paper analyzes the capacity of CP-DSSS in comparison with Orthogonal Frequency-Division Multiplexing (OFDM).  We show that CP-DSSS achieves the same capacity as OFDM and can be optimized with similar precoding methods (e.g., water-filling).  Because of its spread spectrum nature, CP-DSSS can operate as secondary network using the same spectrum as the primary 4G or 5G network, but transmitting at much lower power. Accordingly, the combination of primary and secondary signals in the envisioned setup may be viewed as a power NOMA (non-orthogonal multiple access) technique where primary signals are detected and subtracted from the received signal first before detecting the secondary signals.  In order to operate at a sufficiently low interference level to the primary network, details of CP-DSSS capacity for symbol rate reduction and multi-antenna operation are developed.  The capacity limits established in this paper can be used as a baseline to evaluate the performance of future CP-DSSS receiver architectures for single- and multi-user scenarios.

\end{abstract}

\section{Introduction}
Cyclic prefix direct sequence spread spectrum (CP-DSSS) is a novel waveform that was originally proposed as a solution to ultra-reliable low latency communication (URLLC) \cite{Aminjavaheri:2018:CP_DSSS_Intro,Farhang:2019:Scheduling_URLLC,Stevens:2020}.  These references introduced CP-DSSS as a signaling method for sending a small number of information/control bits within each OFDM symbol interval concurrent with other network communication signals.  

CP-DSSS can also be used as a multi-user data channel to achieve massive machine type communication (mMTC) network density objectives as explored in \cite{Jenkins:2020:IoT_CP_DSSS} and \cite{Kenney:2020:Multi_user_CP_DSSS}.  Unlike URLLC, which is concerned largely with latency, probability of detection, and supporting packet formats, mMTC endeavors to support a massive number of low-cost and low-power terminals.  As low-cost and low-power transceivers are designed, it is important to compare their performance to the theoretical bounds.  This paper presents a capacity analysis of CP-DSSS for a point-to-point link and establishes the peak per-user capacity, assuming an ideal receiver.  Although the developed results are applicable to any signal-to-noise (SNR) ratio, the emphasis of this paper is on the low SNR regime where CP-DSSS as an mMTC technology may be found most useful. The interesting findings here are that (i) across the SNR range, CP-DSSS offers the same capacity as orthogonal frequency division multiplexing (OFDM), including the water-filling capacity; (ii) in the low SNR regime, most of the predicted capacity may be harvested through a matched filter detector, in the uplink, and through a time-reversal precoding, in the downlink \cite{Kenney:2020:Multi_user_CP_DSSS}.

The CP-DSSS waveform uses Zadoff-Chu (ZC) sequences to spread each symbol across the signal bandwidth.  By using a different cyclic shift of the ZC spreading sequence for each symbol, multiple symbols can be multiplexed together in an orthogonal manner.  This allows CP-DSSS to operate up to one symbol per sample before accounting for the extra overhead from the cyclic prefix (CP).  The addition of the CP allows CP-DSSS to fit nicely within the OFDM symbol timing established as part of LTE (Long-Term Evolution) and NR (New Radio) networks.  The CP also preserves circular convolution, which produces convenient mathematical structures that simplify the current analysis as well as future receiver processing.  Another benefit of CP-DSSS is that each data symbol is transmitted through the same channel.  As a result, severe fading or high levels of interference affecting part of the transmission band are seen equally by each symbol, and a special feature of CP-DSSS, [2], allows digitally filtering out the offending portions of the band prior to despreading at the receiver.

Femtocells have been proposed as a means of increasing the density of users in a network for mMTC.  In particular, some researchers have proposed the use of secondary networks that serve users/devices that need lower data rates \cite{Andrews2012}.  In fact, two-tier femtocell networks have been extensively researched \cite{Andrews2012,Chandrasekhar2009,Chandrasekhar2009a,Condoluci2015,Abouei2013} and identified as a 3GPP compliant architecture to reduce radio and network congestion \cite{Dawy2017}.  While CP-DSSS can be used to operate femtocells at a different carrier frequency than the primary network, it can also be used to form a secondary network using the same spectrum as the primary network.  In doing so, CP-DSSS must operate at lower SNR levels in order to reduce interference to the primary network to a negligible level.  The combination of primary and secondary signals in the envisioned setup may be viewed as power NOMA (non-orthogonal multiple access) technique, where primary signals are detected and subtracted from the received signal prior to detecting the secondary signals \cite{Islam:2017:NOMA}.

As discussed in the latter parts of this paper, CP-DSSS offers a trivial symbol rate reduction technique that greatly simplifies the CP-DSSS transceivers without significantly affecting the achievable capacity in the low SNR regime.  Another enabling feature of CP-DSSS that allows for low-power operation is the use of multiple antennas at the femtocell gateway (FGW), which serves as a hub for the femtocell.  The addition of multiple antennas to the FGW also allows for efficient separation of users in a multi-user scenario as is addressed in \cite{Kenney:2020:Multi_user_CP_DSSS}.  The femtocell scenario emphasizes the importance of the capacity analysis in the low SNR regime, especially for cases where symbol rate reduction and multiple FGW antennas are employed.

This paper is organized as follows. A summary of the CP-DSSS waveform is presented in Section II, where we discuss information symbol spreading at the transmitter and despreading at the receiver. This leads to a compact equation of the system model that relates the transmitted data symbols and the received signal after despreading. The system model will be used in Section III to evaluate the capacity of CP-DSSS, both with and without precoding performance optimizations, in comparison to OFDM. In Section IV, we show that precoding of the information symbols by a time reversal of the channel impulse response approaches the water-filling capacity of OFDM. Section V is devoted to some additional CP-DSSS features that enable operation in the low SNR regime in support of femtocells that operate as a secondary network in the same spectrum as the primary network. The concluding remarks of the paper are presented in Section VII.

\section{Waveform Summary}
As discussed in \cite{Aminjavaheri:2018:CP_DSSS_Intro}, the CP-DSSS waveform uses Zadoff-Chu (ZC) sequences to spread symbols across the alotted bandwidth.  A ZC sequence is orthogonal to cyclically shifted versions of the same sequence, so it is possible to transmit up to $N$ symbols simultaneously, where $N$ is the length of the ZC sequence.  Since the intent of this paper is to maximize capacity, it is first assumed that $N$ symbols will be sent per CP-DSSS block.  The possibility of transmitting less than N symbols per CP-DSSS block is discussed in Section V.  After the symbols have been spread with orthogonal versions of the ZC sequence, they are added together to form a CP-DSSS payload block.  The set of ZC sequences is represented as $\mathbf{Z}$ where the columns of $\mathbf{Z}$ are the orthogonal ZC sequences obtained by circularly shifting a root ZC sequence.  The last $M$ samples of the block are repeated at the beginning to form a cyclic prefix.  The length of $M$ is sized to be equal to the maximum delay spread of the channel as in the case for OFDM in a 4G LTE or 5G NR context.  For the purposes of this paper, we assume that $N$ and $M$ are identical to OFDM parameters, where $N$ is the number of OFDM subcarriers.

Demodulation of the CP-DSSS waveform requires the signal to be despread by left-multiplying the received vector by $\mathbf{Z}^{\rm H}$.  The resulting signal takes the form
\begin{align}
\tilde{\mathbf{y}}&= \mathbf{Z}^{\textrm{H}} \mathbf{y} \nonumber \\
\tilde{\mathbf{y}}&= \mathbf{Z}^{\textrm{H}} \left( \mathbf{HZs}+ \mathbf{v} \right) \nonumber \\
\tilde{\mathbf{y}}&=\mathbf{Hs}+\tilde{\mathbf{v}}, \label{eq:1}
\end{align}
where $\mathbf{s}$ is the vector of transmitted symbols, $\mathbf{y}$ is the received signal, $\tilde{\mathbf{v}}$ is the noise vector after being left multiplied by $\mathbf{Z}^{\rm H}$, and $\mathbf{H}$ is an $N \times N$ matrix with a circular structure.  The columns of $\mathbf{H}$ consist of zero-padded versions of the channel impulse response, $\mathbf{h}$, where successive columns are cyclicly shifted down by one, forming a circulant matrix.  Since $\mathbf{Z}^{\rm H}$ is a unitary matrix, the statistics of the channel noise are unchanged (i.e., $\mathbf{ v } \sim \mathcal{N}(0,\sigma_{v}^2 \mathbf{ I })$ and $\tilde{\mathbf{ v }} \sim \mathcal{N}(0,\sigma_{v}^2 \mathbf{ I } )$, where $\mathbf{I}$ is the identity matrix).  The final expression of \eqref{eq:1} results from the fact that $\mathbf{H}$ and $\mathbf{Z}$ are circulant matrices.  Consequently their order is commutable, and $\mathbf{Z}^{\textrm{H}} \mathbf{Z} = \mathbf{I}$.  Based on \eqref{eq:1} and the description of $\mathbf{H}$, it can be seen that the received signal $\mathbf{y}$ is simply the sum of the transmitted symbols after they have been circularly convolved with the channel impulse response plus the channel noise.

\section{Capacity Analysis}
Capacity is defined as the mutual information between transmitted symbols, $\mathbf{s}$, and the received samples after despreading, $\tilde{\mathbf{y}}$.  This can be expressed as
\begin{equation}
I(\tilde{\mathbf{y}};\mathbf{s})=h(\tilde{\mathbf{y}})-h(\tilde{\mathbf{y}}\vert \mathbf{s}), \label{eq:2}
\end{equation}
where $h(\tilde{\mathbf{y}})$ and $h(\tilde{\mathbf{y}}\vert \mathbf{s})$ are the differential entropies.  From Theorem 9.4.1 in \cite{Cover:1991:Info_Theory} we see that the differential entropy of a multivariate normal distribution represented by the vector $\mathbf{x}$ with covariance matrix $\mathbf{C}_{xx}$ is given as
\begin{equation}
h(\mathbf{x})= \log_2( (\pi e)^n\vert \mathbf{C}_{xx}\vert ), \label{eq:3}
\end{equation}
where $n$ is the number of elements of $\mathbf{x}$.  Based on this definition we can define the following for the CP-DSSS signal:
\begin{equation}
h(\tilde{\mathbf{y}})=\log_2( (\pi e)^N\vert \mathbf{C}_{\tilde{y}\tilde{y}}\vert ) \label{eq:4}
\end{equation}
\begin{equation}
h(\tilde{\mathbf{y}}\vert \mathbf{s})= \log_2( (\pi e)^N\vert \mathbf{C}_{\tilde{y}s}\vert ). \label{eq:5}
\end{equation}
Note that because $\mathbf{s}$ is independent of $\tilde{\mathbf{v}}$, we can express $\mathbf{C}_{\tilde{y}\tilde{y}}$ and $\mathbf{C}_{\tilde{y}s}$ as follows:
\begin{equation}
  \mathbf{C}_{\tilde{y}\tilde{y}}=\sigma_s^2\mathbf{H}\mathbf{H}^{\rm H}+\sigma_{v}^2\mathbf{I} \label{eq:6}
\end{equation}
\begin{equation}
\mathbf{C}_{\tilde{y}s}=\sigma_{v}^2\mathbf{I} \label{eq:7}
\end{equation}
where $\mathbf{I}$ is the identity matrix with dimension $N$.

Substituting \eqref{eq:4} and \eqref{eq:5} into \eqref{eq:2} yields the following relation for the mutual information:
\begin{equation}
I(\tilde{\mathbf{y}};\mathbf{s})= \log_2 \left( \frac{\vert \mathbf{C}_{\tilde{y}\tilde{y}}\vert}{\vert \mathbf{C}_{\tilde{y}s}\vert} \right), \label{eq:8}
\end{equation}
which can be further reduced by using \eqref{eq:6} and \eqref{eq:7} to obtain
\begin{equation}
  I(\tilde{\mathbf{y}};\mathbf{s})= \log_2 \left( \abs{ \mathbf{I}+\frac{\sigma_{s}^2}{\sigma_{v}^2}\mathbf{H}\mathbf{H}^{\rm H} } \right). \label{eq:9}
\end{equation}

For clarity, we make the substitution $\mathbf{A}=\mathbf{H}\mathbf{H}^{\rm H}$ and note that $\mathbf{A}$ can be represented as $\mathbf{Q}^{\rm H} \mathbf{\Lambda}_A \mathbf{Q}$, where $\mathbf{Q}$ is a unitary matrix, and $\mathbf{\Lambda}_A$ is a diagonal matrix.  Since $\mathbf{Q}$ is unitary, it holds that $\mathbf{I}=\mathbf{Q}^{\rm H} \mathbf{Q}$, and \eqref{eq:9} can then represented as
\begin{equation}
  I(\tilde{\mathbf{y}};\mathbf{s})= \log_2 \left( \abs{ \mathbf{Q}^{\rm H} \mathbf{Q}+\frac{\sigma_{s}^2}{\sigma_{v}^2}\mathbf{Q}^{\rm H}\mathbf{\Lambda}_A \mathbf{Q} } \right). \label{eq:10}
\end{equation}

It can now be seen that the determinant in \eqref{eq:10} can be factored as $\abs{ \mathbf{Q}^{\rm H} \left( \mathbf{I}+\frac{\sigma_{s}^2}{\sigma_{v}^2}\mathbf{\Lambda}_A \right) \mathbf{Q} }$.  Since the order of $N \times N$ matrices in a determinant are commutable, the determinant can be further reduced to $\abs{ \mathbf{I}+\frac{\sigma_{s}^2}{\sigma_{v}^2}\mathbf{\Lambda}_A }$.  It is now simple to compute the determinant of the resulting diagonal matrix as $\prod_{i=0}^{N-1}{\left( 1+\frac{\sigma_{s}^2}{\sigma_{v}^2}\lambda_{A,i} \right)}$, where $\lambda_{A,i}$ is the $i^{th}$ eigenvalue of the matrix $\mathbf{A}$.  The mutual information can now be represented as 
\begin{equation}
I(\tilde{\mathbf{y}};\mathbf{s})= \sum_{i=0}^{N-1} \log_2 \left( 1+\frac{\sigma_{s}^2}{\sigma_{v}^2}\lambda_{A,i} \right). \label{eq:11} 
\end{equation}

As noted above, the matrix $\mathbf{H}$ is a circulant matrix, so according to Section 8.2.2 of \cite{Farhang:2013:Adaptive_Filters} the matrix can be diagonalized by a Discrete Fourier Transform (DFT) matrix and the diagonal elements are the eigenvalues of the circulant matrix.  Since $\mathbf{H}$ is circulant, $\mathbf{H}^{\rm H}$ is also circulant, and the resulting eigenvalues are the conjugate of the eigenvalues of $\mathbf{H}$.  Another property of a circulant matrix is that the eigenvalues are obtained by taking the DFT of the first column of the matrix.  Since the first column of $\mathbf{H}$ is the impulse response $\mathbf{h}$ that has been zero-padded to a length $N$, the eigenvalues of $\mathbf{H}$ are simply the values of the $N$-point DFT of $\mathbf{h}$ (i.e., $\lambda_{H,i}=H_i$).  It can be shown that for the matrix $\mathbf{A} = \mathbf{H}\mathbf{H}^{\rm H}$, the eigenvalues are the magnitude squared of the $N$-point DFT of $\mathbf{h}$ (i.e., $\lambda_{A,i}=\vert \lambda_{H,i} \vert ^2=\vert H_i \vert ^2$).  Hence for a bandwidth of $NW$ (e.g., $N=2048$ and $W=15$kHz), the CP-DSSS capacity without precoding (i.e., equal power or EP) can be expressed as
\begin{equation}
C_{\textrm{ EP } } =\sum_{i=0}^{N-1} W\log_2 \left( 1+\frac{\sigma_{s}^2}{\sigma_{v}^2}\vert H_i \vert ^2 \right). \label{eq:12}
\end{equation}

\subsection{Capacity Comparison with OFDM}
The capacity of OFDM with equal power on each subcarrier is achieved by optimizing the rate for each of the $N$ subcarriers.  The channel path gain for subcarrier $i$ is $\vert H_i \vert ^2$.  The signal power is $\sigma_s^2$ and the noise power is $\sigma_v^2$.  The resulting capacity for OFDM with equal power (EP) for all subcarriers (see Section 4.3 of \cite{Goldsmith:2005:Wireless_Comms}) can also be expressed as in \eqref{eq:12}.  This assumes that each subcarrier is optimized in rate for the OFDM case.  Individualized rate optimization is not a restriction for CP-DSSS capacity, where the rate of all the symbols is identical.  Optimizing rate for each subcarrier presents a problem for practical systems because the effective rate is a combination of modulation format and forward error correction (FEC) coding.  Customizing modulation and coding for an individual subcarrier could result in a FEC codeword duration that exceeds the coherence time of the channel.  Consequently, today's systems aggregate subcarriers over a time period in a resource block and assign the same modulation and coding rate to all subcarriers in the block, resulting in some sacrifice of capacity for OFDM.

It is well known that the capacity of OFDM can be further optimized through the water-filling (WF) algorithm.  If more power is allocated to subcarriers with better channels, then capacity is improved.  The capacity for frequency-selective fading channels with block frequency-selective fading (e.g., OFDM-channelization) is given in \cite{Goldsmith:2005:Wireless_Comms} to be
\begin{equation}
C_{\textrm{ WF } }=\sum_{i=0}^{N-1} W\log_2(1+\frac{P_i}{\sigma_{v}^2}\vert H_i \vert ^2 ), \label{eq:14}
\end{equation}
where $P_i$ is the power allocated to each channel, and $\sum_i{P_i}=P_{\rm total}$.  The value of $P_i$ is specified with the following relation:
\begin{equation}
\frac{P_i}{P_{\rm total}}=\begin{cases} 1/\gamma_0 - 1/\gamma_i &\gamma_i \geq \gamma_0, \\ 0 &\gamma_i < \gamma_0, \end{cases} \label{eq:15}
\end{equation}
for the threshold $\gamma_0$ and for $\gamma_i=P_{\rm total} \vert H_i \vert ^2 / \sigma_v^2$.

\subsection{Optimized CP-DSSS Capacity via Precoding}
Even though CP-DSSS does not have a host of subcarriers to optimize like OFDM, the modulation format still contains a significant amount of flexibility when it comes to optimizing performance to improve capacity.  One advantage of CP-DSSS over OFDM is that each data symbol in the CP-DSSS symbol is spread over the entire bandwidth with a ZC sequence.  As a result, each data symbol experiences the highs and lows of the frequency-selective channel.  In order to optimize CP-DSSS, a precoding matrix, $\mathbf{G}$, is applied to the data vector, $\mathbf{s}$, prior to  spreading with the ZC sequences.  The resulting received vector (after despreading) now becomes
\begin{equation}
\tilde{\mathbf{y}}=\mathbf{HGs}+\tilde{\mathbf{v}}. \label{eq:16}
\end{equation}

The selection of $\mathbf{G}$ should be the solution of the following optimization problem based on \eqref{eq:9}:
\begin{maxi}
  {\mathbf{G}}{\left( \vert \mathbf{I}+\frac{\sigma_{s}^2}{\sigma_{v}^2}\mathbf{HG} \mathbf{G}^{\rm H} \mathbf{H}^{\rm H} \vert \right)}{}{}
  \addConstraint{{\rm tr}(\mathbf{G}^{\rm H}\mathbf{G})=P_{\rm total},}
  \label{eq:17}
\end{maxi}
where it is assumed that $\sigma_s^2=1$ (i.e., unity symbol power constraint).  The optimum selection of $\mathbf{G}$ is an open-ended problem with no clear path for a solution.  However, if $\mathbf{G}$ were constrained to be a circulant matrix like $\mathbf{H}$, then the optimization problem can be simplified.  Given that a circulant matrix can be diagonalized by the DFT matrix, we can express the diagonal elements of $\mathbf{B}=\mathbf{G} \mathbf{G}^{\rm H}$ as $\lambda_{B,i}$, and the optimization problem is reduced to
\begin{maxi}
  {\lambda_{B,i}}{\left(\prod_{i=0}^{N-1} \left(1+\frac{\sigma_{s}^2}{\sigma_{v}^2}\lambda_{A,i}\lambda_{B,i} \right)\right)}{}{}
  \addConstraint{\sum_i{\lambda_{B,i}} \leq P_{\rm total}.}
  \label{eq:18}
\end{maxi}

It was previously shown that $\lambda_{A,i}=\vert H_i \vert ^2$.  Clearly, if $\sigma_s^2 \lambda_{B,i}$ were set equal to $P_i$ in \eqref{eq:15}, then the CP-DSSS capacity would equal the OFDM capacity with WF shown in \eqref{eq:14}.  Assuming that $\sigma_s^2=1$, the process to find $\mathbf{g}$, the first column of the circulant matrix $\mathbf{G}$, is to first solve for the eigenvalues of $\mathbf{G}$ (i.e $\lambda_{G,i}$), which can simply be the square root of the elements $P_i$.  Note that since the intent is to provide power scaling over frequency, the phase of $\lambda_{G,i}$ is not important.  The vector $\mathbf{g}$ is then found by taking the inverse DFT of $\lambda_{G,i}$.  Using cyclical shifts of $\mathbf{g}$ to form the circulant matrix $\mathbf{G}$, the CP-DSSS capacity with prescaling now equals the OFDM capacity with WF.  To verify that the power constraint was maintained, we show that $\text{trace}(\mathbf{G}^{\rm H}\mathbf{G})= \sum_i \lambda_{G,i}^2 = \sum_i \lambda_{B,i} = \sum_i P_i = P_{\rm total}$.

\section{Suboptimal Precoding}
The process of calculating the optimal precoding matrix that achieves the Shannon capacity requires that the WF algorithm be executed. The WF algorithm, as noted above, identifies the eigenvalues of the circulant precoding matrix $\mathbf{G}$ according to the constrained maximization procedure in \eqref{eq:18}, and then applies an inverse DFT step to obtain the first column of $\mathbf{G}$. The WF algorithm, despite leading to an optimum design, is computationally expensive to implement. Hence, any suboptimum precoder that reduces the complexity but still leads to a good result will be valuable. Here, we propose one such precoder. To this end, we note that the WF algorithm has the effect of emphasizing portions of the spectrum with low loss and deemphasizing (or even not using) portions of the spectrum with high loss. At the same time, we note that using a time-reserved version of the channel impulse response as the precoder will have a similar effect, although it will not zero-out portions of the spectrum as done in WF. The time reversal (TR) precoder is defined as the Hermitian of the circulant channel matrix, $\mathbf{H}$, after scaling for unity power, which can be defined as
\begin{equation}
  \mathbf{G}= \frac{\mathbf{H}^{\rm H}}{\sqrt{{\rm tr}(\mathbf{H}^{\rm H}\mathbf{H})/N}}, \label{eq:19}
\end{equation}
where ${\rm tr}()$ is the trace operator.  The trace can alternatively be calculated by taking the sum of the eigenvalues, which are readily available (i.e., ${\rm tr}(\mathbf{H}^{\rm H}\mathbf{H})={\rm tr}(\mathbf{A})=\sum_{i=0}^{N-1}{\lambda_{A,i}}=\sum_{i=0}^{N-1}{\vert H_i \vert ^2}$).

\begin{figure}[!t]
\centering
\includegraphics[width=3.6in, clip=true, trim=4cm 8.5cm 4cm 8.5cm ]{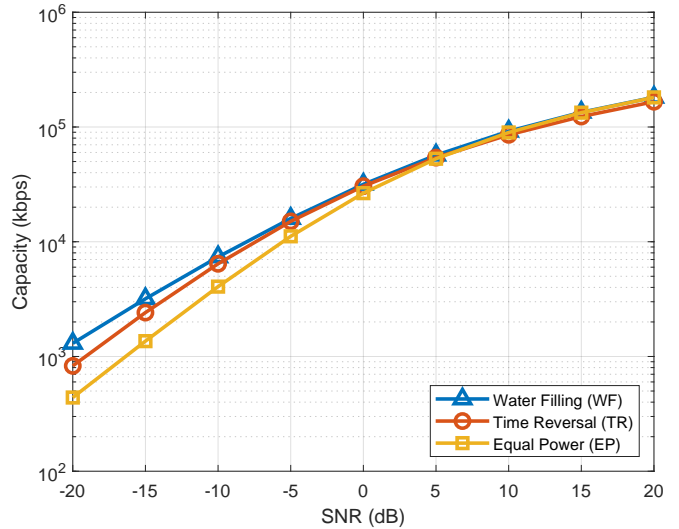}
\caption{Comparison of single-user CP-DSSS capacity with WF and TR precoding as well as EP (no precoding).  The curves represent the average theoretical capacity (i.e., no detector specified) taken over many randomly generated channels.  Each channel is 130 samples long with an exponential roll-off factor of 25. }
\label{Capacity_comparison}
\end{figure}

The average capacity of the TR precoder along with that of the WF precoder and the EP are presented in Fig. \ref{Capacity_comparison}.  The channel used here is $130$ samples long with an exponential roll-off factor of $25$, and the presented results are obtained by averaging over $1000$ realizations of this channel.  For the low SNR regime, WF and TR perform better than EP. As SNR increases, the EP capacity converges to WF capacity, which
is expected given the relation in \eqref{eq:15}.  As SNR increases, the $\gamma_i \geq \gamma_0$ condition prevails and the $1/\gamma_i$ term approaches zero, leaving an asymptotic power level of $P_{\rm total}/\gamma_0$ for each spectral bin (i.e., equal power). TR capacity is slightly worse than WF and EP for higher SNR, and this capacity gap remains for a wide range of SNR values. Only when SNR exceeds several hundreds of decibels (not shown here) the TR performance asymptotically approaches that of WF and EP.  

We consider the latter deviation of TR precoding from the WF precoding as not relevant to the specific applications where the CP-DSSS may be applied. As discussed in the next section, CP-DSSS, being a spread spectrum technique, may be found to be most useful in applications where it operates under the noise floor. Hence, we will be mostly interested in low SNR regimes where TR precoding provides a great performance improvement over EP transmission.

\section{Additional CP-DSSS Features}
CP-DSSS is a highly flexible waveform which may be adapted to operate as a secondary network. It can operate with a wide range of spreading gains, allowing the power spectral density to be reduced in order to avoid interference with the primary network. Thus far, this paper has focused on achieving the maximum capacity of the CP-DSSS waveform, where the spreading gain was set to unity; the symbol rate was set to be only slightly below the transmission bandwidth. In order to operate as a secondary network, the spreading gain may be increased, i.e., the symbol rate is decreased compared to the transmission bandwidth, in order to reduce the power spectral density of the CP-DSSS waveform to a level of low interference to the primary network. In the following subsection, we present the mathematical equations that formalize this lower symbol rate scenario and evaluate the channel capacity in the low SNR regime to quantify the impact of such reduction in symbol rate on the achievable data rate. The interesting finding here is that reduction of symbol rate, over a wide range, has a small impact on the channel capacity.

\subsection{Symbol Rate Reduction} \label{sec:symbol_rate_reduction}

\begin{figure}[!t]
\centering
\includegraphics[width=3.6in, clip=true, trim=4cm 8.5cm 4cm 8.5cm ]{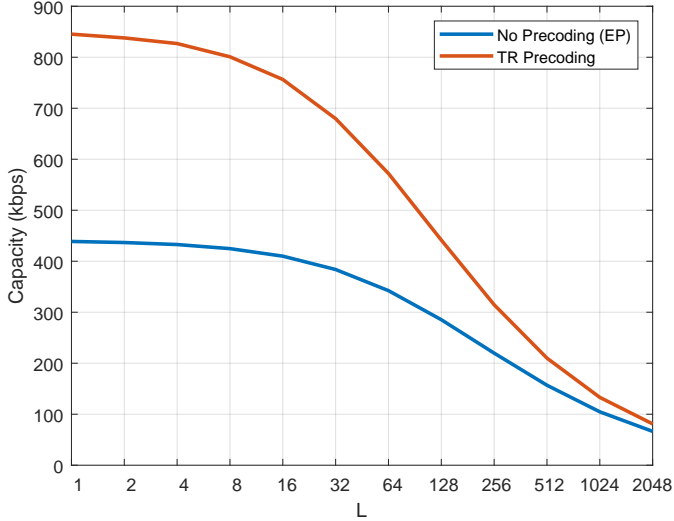}
\caption{Single-user CP-DSSS capacity vs. L for 2048 potential payload symbols at 15 kHz subcarrier spacing at SNR = -20 dB.  The curves represent average capacity taken over many randomly generated channels.  Each channel is 130 samples long with an exponential roll-off factor of 25. }
\label{Capacity_vs_L1}
\end{figure}

Symbol rate reduction is accomplished by forming an expander matrix, $\mathbf{E}_L$, which is based on the symbol reduction factor, $L$.  The form of $\mathbf{E}_L$ can be described as an identity matrix of dimension $N/L$ (i.e., $\mathbf{I}_{N/L}$) that has been upsampled in the vertical dimension by a factor of $L$.  In other words, after each row of $\mathbf{I}_{N/L}$, $L-1$ rows of zeros are inserted, resulting in an $N \times N/L$ matrix.  When symbol reduction and precoding are employed, the despread received signal and corresponding capacity take the form of
\begin{equation}
\tilde{\mathbf{y}}=\mathbf{HGE}_L \mathbf{s}+\tilde{\mathbf{v}}, \label{eq:20}
\end{equation}  
\begin{equation}
C=W \textrm{log}_2 \left( \lvert \mathbf{I} + L \frac{\sigma_s^2}{\sigma_v^2} \mathbf{HGE}_L\mathbf{E}_L^{\textrm{H}}\mathbf{G}^{\textrm{H}}\mathbf{H}^{\textrm{H}} \rvert \right), \label{eq:20b}
\end{equation}
where $\mathbf{s}$ has $N/L$ symbols and each symbol is scaled by $\sqrt{L}$ such that the power per symbol is $L$ times $\sigma_s^2$.

Next, to reveal some interesting properties of the CP-DSSS waveform in the application of secondary networks, we use the channel model in \eqref{eq:20} to evaluate the capacity in low SNR regimes as a function of $L$.  Fig. \ref{Capacity_vs_L1} shows the capacity of CP-DSSS with varying amounts of symbol rate reduction at an SNR of $-20$ dB for the same channel characteristics used in Section IV.  As in Fig. \ref{Capacity_comparison}, the TR precoding outperforms the EP case by a large margin.  In this low SNR regime where a secondary network would operate, the capacity falls off slowly with $L$ until $L$ becomes large.  Using a value of $L>1$ reduces the complexity of the decoder for achieving capacity.  For example, to achieve the capacity of $845$ kbps for $L=1$, a code rate of $(845~\mbox{kbps})/(2048\times 15~\mbox{kHz})=0.0275$ bits per symbol would be needed.  In comparison, the capacity is only $10\%$ lower at $756$ kbps for $L=16$, and a code rate of $(756~\mbox{kbps})/(128\times 15~\mbox{kHz})=0.394$ bits per symbol would be needed, which is much less complex to decode, making higher values of $L$ very attractive as long as the capacity is not reduced significantly.

\subsection{Massive MIMO Scenario} 
Another important aspect of CP-DSSS that greatly enhances its applicability to a secondary network is that the waveform is easily adaptable to a deployment with multiple antennas at the FGW.  Multiple antennas at the FGW provide a two-fold benefit\textemdash (i) more users can be supported due to spatial decorrelation and (ii) transmit power levels can be reduced due to antenna array gain. With the addition of more antennas on the FGW comes additional processing. However, no additional processing is needed by the secondary network terminals, which is consistent with the desire for low-complexity and low-power devices for mMTC. When the number of FGW antennas grows large, the femtocell can benefit from the effect of massive MIMO, where simple linear detectors have performance approaching capacity \cite{Hoydis:2013}. Here, we present the mathematical equations that explain the channel model when there are more than one antenna at the FGW and study the impact of increasing the number of FGW antennas on the system capacity.

In the uplink, the channel model in \eqref{eq:1} is modified with the following substitutions to facilitate multiple antennas at the FGW:
\begin{equation}
\tilde{\mathbf{y}}=
  \begin{bmatrix}
    \tilde{\mathbf{y}}^{(1)} \\
    \tilde{\mathbf{y}}^{(2)} \\
    \vdots \\
    \tilde{\mathbf{y}}^{(M)}
  \end{bmatrix}, \\
\mathbf{H}=
  \begin{bmatrix}
    \mathbf{H}^{(1)} \\
    \mathbf{H}^{(2)} \\
    \vdots \\
    \mathbf{H}^{(M)}
  \end{bmatrix}, \\
\tilde{\mathbf{v}}=
  \begin{bmatrix}
    \tilde{\mathbf{v}}^{(1)} \\
    \tilde{\mathbf{v}}^{(2)} \\
    \vdots \\
    \tilde{\mathbf{v}}^{(M)}
  \end{bmatrix},  
 \label{eq:22}
\end{equation}
where $M$ is the number of FGW antennas, $\tilde{\mathbf{y}}^{(1)}$ through $\tilde{\mathbf{y}}^{(M)}$ are received signal vectors at the specified antennas, $\mathbf{H}^{(1)}$ through $\mathbf{H}^{(M)}$ are the circulant channel matrices corresponding to each antenna, and $\tilde{\mathbf{v}}^{(1)}$ through $\tilde{\mathbf{v}}^{(M)}$ are the noise vectors for each antenna.  In the uplink scenario, the terminal does not precode the data, so there is no matrix $\mathbf{G}$.

\begin{figure}[!t]
\centering
\includegraphics[width=3.6in, clip=true, trim=4cm 8.5cm 4cm 8.5cm ]{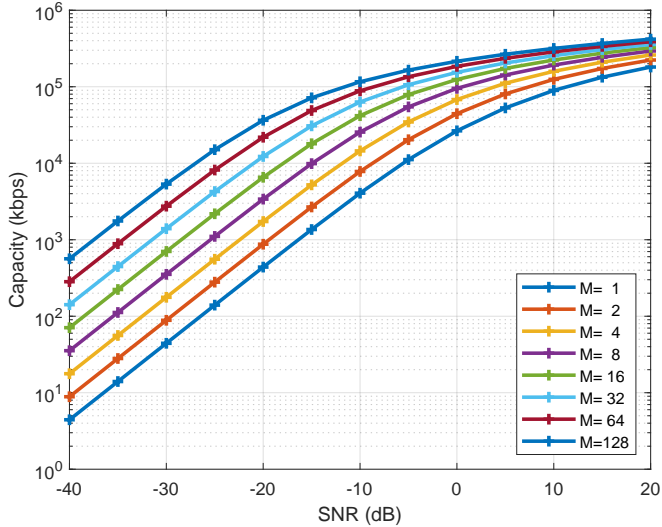}
\caption{Single-user CP-DSSS capacity without precoding for an uplink scenario with a multi-antenna FGW.  The curves represent average capacity taken over many randomly generated channels.  Each channel is 130 samples long with an exponential roll-off factor of 25. }
\label{Capacity_multi_antenna_UL}
\end{figure}

When there are multiple  FGW antennas, the matrix $\mathbf{H}$ in \eqref{eq:22} is no longer circulant, but the capacity can still be calculated based on \eqref{eq:11}.  Fig. \ref{Capacity_multi_antenna_UL} shows how the capacity without precoding improves for multiple channels, where the channels seen by each antenna are selected i.i.d. with length of 130 samples and an exponential roll-off factor of 25 samples.  The curve for $M=1$ in Fig. \ref{Capacity_multi_antenna_UL} is identical to the EP case in Fig. \ref{Capacity_comparison}.  Increasing $M$ further provides higher capacity for each SNR (i.e., approximately $3$ dB for each doubline of $M$ at low SNR value), which allows the secondary network terminal to transmit at a lower power than it otherwise would for a single-antenna FGW.  This permits more secondary network terminals to participate simultaneously without increasing the interference to the primary network.

To accommodate the downlink scenario, precoding is necessary at the transmitter and the following substitutions are made into the channel model in \eqref{eq:16}:
\begin{equation}
\mathbf{H} = 
	\begingroup 
		\setlength\arraycolsep{4pt}
		\begin{bmatrix}
		\mathbf{H}^{(1)} & \mathbf{H}^{(2)} & \dots & \mathbf{H}^{(M)}
		\end{bmatrix},
	\endgroup  
\mathbf{G}=
	\begin{bmatrix}
    	\mathbf{G}^{(1)} \\
	    \mathbf{G}^{(2)} \\
	    \vdots \\
	    \mathbf{G}^{(M)}
	\end{bmatrix}, \\
\label{eq:23}
\end{equation}
where $M$ is the number of FGW antennas, $\mathbf{H}^{(1)}$ through $\mathbf{H}^{(M)}$ are the circulant channel matrices, and $\mathbf{G}^{(1)}$ through $\mathbf{G}^{(M)}$ are the precoding matrices corresponding to each antenna element.  The objective in precoding for the downlink is to maximize the signal at the intended receiver and minimize interference.  The downlink model in this paper shows the formulation for a single receiver, but it could be easily extended for the intended multiple user scenario, where interference between users must be addressed.  In order to maintain the same transmit signal power as in the single antenna case, $\mathbf{G}$ must be selected such that $\textrm{tr}(\mathbf{GG}^{\textrm H})=N$.  One of the results reported in \cite{Hoydis:2013} is that as the number of antennas increase, precoding can be achieved with simple linear operations such as TR.  Hence, we can construct the constituent precoding matrices of $\mathbf{G}$ as
\begin{equation}
  \mathbf{G}^{(j)}= \frac{\mathbf{H}^{(j)\rm H}}{\sqrt{{\rm tr}(\mathbf{H}^{(j)\rm H}\mathbf{H}^{(j)})\frac{M}{N}}}, \label{eq:24}
\end{equation}
where $j$ is the FGW antenna index, and the extra factor of $\sqrt{M}$ in the denominator achieves the constraint $\textrm{tr}(\mathbf{GG}^{\textrm H})=N$.  
The capacity results for the DL case are very similar to Fig. \ref{Capacity_multi_antenna_UL} for larger values of $M$.  For smaller values of $M$ (e.g., $M < 32$), the DL theoretical capacity is higher than the UL capacity for SNR values less than $0$ dB due to the precoding, but the precoding gains phase out as $M$ increases.  While the results shown here for uplink and downlink are for cases where $L=1$, higher values of $L$ can be used as described in Section \ref{sec:symbol_rate_reduction} to reduce complexity and still perform close to the $L=1$ capacity.

\section{Conclusion}
CP-DSSS is a novel direct sequence spread spectrum waveform that has recently been proposed for low-rate data transmission in wireless networks. Because CP-DSSS can be configured to have the same number of samples as an OFDM symbol,  CP-DSSS can take advantage of the synchronization signals in the LTE or 5G NR network.  Furthermore, by transmitting at lower power and cancelling the OFDM signal of the primary network in a power-domain NOMA fashion, CP-DSSS can co-exist as a secondary network with negligible interference to the primary network.

This paper presented an in-depth analysis of the capacity of CP-DSSS in various channel conditions. Capacity formulae with and without precoding were presented, and it was shown that CP-DSSS can match the capacity of OFDM. With precoding, in particular, CP-DSSS offers the same capacity as OFDM when the water-filling power allocation algorithm is applied. We also showed that a good portion of this capacity (particularly in low SNR regimes) can be harvested by using a simple precoder that is based on the time-reverse of the channel impulse response.

The latter part of this paper examined two CP-DSSS features that enable operation in the low SNR regime (e.g., $< -20$ dB). We showed that the CP-DSSS symbol rate can be dramatically reduced with an accompanying reduction in complexity, without sacrificing much capacity. Capacity curves were presented for the uplink scenario without precoding, showing an array gain approximately equal to the number of antennas when operating in the low SNR regime. Finally, the use of CP-DSSS to build a secondary network of femtocells that co-exist with primary networks can facilitate many use cases of device communications beyond 5G.


\end{document}